# Advancing Understanding of Long COVID Pathophysiology Through Quantum Walk-Based Network Analysis

Jaesub Park[1,2,3†], Woochang Hwang[4†], Seokjun Lee[1,2,3‡], Hyun Chang Lee[1,2,3‡], Méabh MacMahon[4], Matthias Zilbauer[1,5,6] and Namshik Han[1,2,3*]

1 Cambridge Stem Cell Institute, University of Cambridge, Cambridge, UK

2 Milner Therapeutics Institute, University of Cambridge, Cambridge, UK

3 Cambridge Centre for AI in Medicine, Department of Applied Mathematics and Theoretical Physics, University of Cambridge, Cambridge, UK

4 Cardiatec Biosciences Ltd, Cambridge, UK

5 Department of Paediatrics, University of Cambridge, Cambridge, UK

6 Department of Paediatric Gastroenterology, Hepatology and Nutrition, Cambridge University Hospitals (CUH), Addenbrooke's, Cambridge, UK

†These authors contributed equally.

‡These authors contributed equally.

*Corresponding author. Email: nh417@cam.ac.uk.

## Abstracts

Long COVID is a multisystem condition characterized by persistent symptoms such as fatigue, cognitive impairment, and systemic inflammation, following COVID-19 infection, yet its mechanisms remain poorly understood. In this study, we applied quantum walk (QW), a computational approach leveraging quantum interference, to explore large-scale SARS-CoV-2–induced protein (SIP) networks. Compared to the conventional random walk with restart (RWR) method, QW demonstrated superior capacity to traverse deeper regions of the network, uncovering proteins and pathways implicated in Long COVID. Key findings include mitochondrial dysfunction, thromboinflammatory responses, and neuronal inflammation as central mechanisms. QW uniquely identified the CDGSH iron-sulfur domain-containing protein family and VDAC1, a mitochondrial calcium transporter, as critical regulators of these processes. VDAC1 emerged as a potential biomarker and therapeutic target, supported by FDA-approved compounds such as cannabidiol. These findings highlight QW as a powerful tool for elucidating complex biological systems and identifying novel therapeutic targets for conditions like Long COVID.

## Introduction

Long COVID, also known as post-acute sequelae of SARS-CoV-2 infection, represents a complex constellation of persistent symptoms that continue or develop weeks to months after the initial COVID-19 infection [1,2]. One cohort study (n=1733) reports at least one long term symptom in 68% of COVID-19 survivors, with symptoms ranging from fatigue and cognitive dysfunction to respiratory problems and cardiovascular complications [3]. The heterogeneous nature of Long COVID poses a substantial challenge to healthcare systems worldwide, as patients can experience vastly different combinations of symptoms with varying degrees of severity [4]. The treatment of Long COVID presents several challenging aspects for healthcare providers. First, the multisystem involvement hinders the development of standardized treatment protocols [5]. Second, the underlying pathophysiological mechanisms remain poorly understood, complicating targeted therapeutic approaches [6]. Additionally, the symptoms often fluctuate over time and vary significantly among patients, making it challenging to assess treatment efficacy [7]. The lack of specific diagnostic markers further compounds these difficulties, as clinicians must rely heavily on patient-reported symptoms and clinical observations [8].

Current computational approaches to understanding and treating Long COVID, while promising, face limitations [9]. For instance, traditional data analysis methods struggle to capture the complex interactions between various symptoms and biological pathways involved in the condition [10]. Moreover, existing computational models often fail to account for the temporal dynamics of symptom progression and the influence of individual patient factors [11]. These limitations highlight the pressing need for innovative computational approaches that can integrate diverse data types, including molecular profiles, clinical observations, and patient-reported outcomes, to uncover hidden patterns within biological networks [12]. Quantum-inspired methods, such as quantum walk (QW) algorithms, which simulate quantum mechanics principles for analyzing network propagation, have demonstrated significant potential in addressing complex graph-based problems, offering a promising avenue to overcome these challenges [13].

In recent years, quantum algorithms, including QW, have shown great potential in efficiently navigating graph-based systems, offering insights that are challenging to obtain through classical methods. QW algorithms extend the concept of classical random walks by leveraging quantum mechanics principles such as superposition and interference. Superposition allows the walker to exist in multiple states simultaneously, while interference enhances its ability to navigate complex graphs by amplifying certain paths and suppressing others [14]. This unique behaviour makes QW especially suited for uncovering patterns in large, interconnected networks. Among the algorithms simulating QW, discrete-time quantum walk (DTQWs) are particularly notable. These walks operate under unitary evolution and consist of two main components: a coin operator, which dictates the direction of the walker, and a shift operator, which updates its position on the graph. This stepwise evolution creates interference patterns, absent in classical random walk with restart (RWR), which may lead to fundamentally different outcomes in network analysis.

Building on the foundational principles of QW, recent research has explored its application in biological systems, demonstrating its ability to address challenges in network analysis and disease prioritization. The distinctive features of QW make it highly effective for uncovering the structural and dynamic properties of real-world networks. For instance, QW has been utilized to rank network nodes and reveal community structures in complex networks [15-17]. More recently, QW has been applied to tackle complex biological problems such as protein–protein interaction prediction, protein–DNA interaction search, and disease gene prioritization [18-20]. These examples highlight the potential of QW, positioning it as a promising tool in modern computational biology and suggesting that it could play a pivotal role in advancing the analysis of complex biological systems.

In this study, we employed quantum-inspired approaches to analyze protein interaction networks associated with Long COVID, focusing on their ability to uncover critical biological mechanisms. Specifically, we utilized the QW algorithm to prioritize experimentally validated proteins linked to Long COVID, demonstrating its superior capability in navigating the complex topology of biological networks compared to traditional method RWR. Our results revealed that QW not only outperformed conventional methods in prioritizing validated Long COVID-associated proteins but also uncovered pathways closely tied to the disease's pathophysiology, including mitochondrial dysfunction. Notably, QW extended its exploration into deeper regions of the network, uncovering critical biological interactions that underlie mitochondrial dysfunction—a hallmark of Long COVID. These findings highlight the potential of QW-based approaches in advancing our understanding of complex diseases such as Long COVID. By overcoming the limitations of traditional network analysis methods, QW offers a robust platform for uncovering novel disease mechanisms and facilitating the identification of therapeutic targets.

# Result

**The SARS-CoV-2–induced protein network reveals the importance of deep exploration**

To elucidate the mechanisms underlying Long COVID, we investigated how SARS-CoV-2 impacts host proteins and pathways using a SARS-CoV-2–induced protein (SIP) network [21-22]. The rationale for constructing the SIP network lies in the assumption based on cause and effect, that proteins significantly affected during acute COVID-19 could contribute to the long-term effects observed in Long COVID.

The SIP network was constructed by integrating Directly Interacting Proteins (DIPs), which directly interact with the SARS-CoV-2 protein, and Differentially Expressed Proteins (DEPs), the latter identified from proteins exhibiting differential expression during acute COVID-19. To provide comprehensive framework for exploring these integrations, the SIP network was designed with three distinct layers: the DIP layer, the DEP layer, and an intermediate hidden layer that bridges the two, consisting of 11,247 nodes and 179,554 unique edges. All shortest paths between DIPs and DEPs were identified, ensuring that all potential interactions were captured within the network.

To assess the impact of SARS-CoV-2 on the SIP network, we applied network propagation methods, including Quantum Walk (QW) and Random Walk with Restart (RWR) (Figure1A). These methods allowed us to model how the viral signal influences proteins across the network. For further analysis, we categorized the network nodes into shallow (distance ≤ 2) and deep (distance > 2) regions, where distance refers to the distance from SARS-CoV-2, enabling us to explore the distinct roles of proximal and distal proteins.

Our analysis revealed distinct patterns in the distribution of proteins across the SIP network, as shown in Figure 1a. Among all nodes (Figure 1B), the majority of proteins in the SIP network were classified as deep nodes, with 50.8% at distance 3, 3.2% at distance 4, and 0.1% at distance 5. Shallow nodes accounted for 45.9%, with 2.9% at distance 1 and 43.0% at distance 2. When focusing on DEPs (Figure 1C), a higher proportion of DEPs were observed in deep regions compared to the general background distribution. Specifically, 61.7% of DEPs were located in deep nodes, with 54.1% at distance 3, 7.1% at distance 4, and 0.5% at distance 5, while only 38.3% are in shallow nodes, with 1.7% at distance 1 and 36.6% at distance 2.

This enrichment of DEPs, which we know are changing in response to COVID-19 infection, in deeper regions suggests that proteins further from the SARS-CoV-2 signal play an important role in mediating the broader effects of the virus. These findings highlight the need to investigate both shallow and deep regions of the SIP network to fully capture the direct and indirect impacts of SARS-CoV-2.

**QW enables comprehensive exploration of the SIP network beyond RWR's limitations**

Proteins that are highly connected and involved in multiple pathways within a protein interaction network may exert significant influence on the network and perform essential functional roles. To identify key proteins of COVID-19, we applied QW and RWR to constructed SIP network. By designating the SARS-CoV-2 node as the starting point for both RWR and QW, we aimed to prioritize nodes frequently visited by walkers along the numerous paths spanning from DIPs to DEPs. This approach is intended to uncover proteins central to COVID-19 that may significantly influence disease progression beyond the acute phase and contribute to the development of Long COVID.

Both RWR and QW algorithms exhibit dynamic fluctuation in node probabilities throughout the iterative process, along with the lack of an inherent guarantee of convergence. Therefore, it is crucial to calculate the mixing time, which is the point at which the node probability distribution reaches a stable state, allowing the determination of the final probability distribution [14]. To address this, we quantified the change in node probabilities at each iteration by calculating the Euclidean distance of probability distribution (EDP) between consecutive iterations. For the RWR, EDP progressively decreased across successive steps, converging to a negligible value ($1.0 \times 10^{-4}$) within a finite number of iterations (Supplementary Figure A). Based on this observation, the mixing time was defined as the iteration at which the EDP reached $1.0 \times 10^{-4}$. The probability distribution at the mixing time was used as the result of the RWR algorithm.

In contrast, the QW exhibited a decrease in the EDP initially but failed to converge to a sufficiently small value, instead oscillating around a constant value (Figure 2A, 2B). To resolve this issue, we identified an approach that calculates an averaged probability distribution and defines the average mixing time [14]. As the iterations progressed in the QW, the Euclidean distance of averaged probability distributions (EDAP) between consecutive iterations gradually stabilized, eventually reaching below $1.0 \times 10^{-4}$ (Figure 2C). The averaged mixing time for QW was defined as the iteration where the change in EDAP fell below $1.0 \times 10^{-4}$. The average probability distribution at the mixing time was used for subsequent analysis of QW results.

After obtaining the results from each algorithm, we compared the characteristics of the walkers in QW and RWR by examining the distribution of probability values for each node in the SIP network based on their distance from the SARS-CoV-2 node (Figure 2D). In the case of QW, the probability distribution for nodes at distances 1 to 4 from the SARS-CoV-2 node appeared nearly uniform, with only a slight decrease at distance 5. In contrast, RWR exhibited a clear decrease in probability values as the distance from the SARS-CoV-2 node increased. This observation is consistent with the well-documented limitation of RWR in effectively exploring nodes located farther from the starting point in large-scale networks. In contrast, the inherent properties of QW appear to overcome this limitation of RWR, enabling effective exploration of the deeper regions of the network.

To further investigate the preceding results, we analyzed the distances from the SARS-CoV-2 node for the top 947 key proteins identified by each algorithm (Figure 2E, see the Methods). Most top nodes identified by QW were at distances of 2 or 3 from the SARS-CoV-2 node, with a smaller fraction at distance 4. Conversely, top nodes identified by RWR were primarily at distances of 1 or 2, with very few nodes at distance 3. These findings suggest that RWR outcomes are strongly distance-dependent, prioritizing proteins closer to the starting node. By contrast, QW assigns relatively uniform probabilities to nodes farther from the starting node, identifying them as key proteins as well. Given that the SIP network represents a collection of paths from SARS-CoV-2 nodes to DEPs, and DEPs exist as far as distance 5, these results suggest that QW sufficiently explores the network more comprehensively and may offer a more accurate assessment of node importance within these pathways.

**QW-identified key proteins show spatial similarity to LCPs in the SIP network**

To evaluate the potential of QW in identifying proteins from the SIP network, which go on to be associated with Long COVID, high-confidence Long COVID proteins (LCPs) were curated from the literature. A recent study by Cervia-Hasler et al. conducted a comprehensive 12-month longitudinal proteomics analysis to investigate changes in blood serum proteins in Long COVID patients (Figure 3A) [23]. Long COVID was defined as the persistence of one or more COVID-19–related symptoms (e.g., fatigue, pulmonary symptoms, gastrointestinal symptoms, cognitive symptoms, etc.) at six months post-infection, without alternative diagnosis. Among the 6,596 human proteins analyzed, 514 differentially expressed proteins, referred to as LCPs, were identified at the 6-month follow-up by comparing Long COVID patients (n = 40) to recovered patients (n = 73).

To determine the spatial distribution of LCPs within the SIP network, 423 of the 514 LCPs were mapped onto the network and categorized by their distances from the starting SARS-CoV-2 node (Figure 3B). Most LCPs were positioned at distances 2 and 3 from the SARS-CoV-2 node, with 60% classified as deep LCPs (distance > 2) and the remaining 40% as shallow LCPs (distance ≤ 2). This distribution suggests that most LCPs reside in regions of the network further from DIPs of SARS-CoV-2.

To explore the functional roles of LCPs in shallow versus deep nodes, we performed overrepresentation analysis (ORA) on the two groups separately using REACTOME database (Figure 3C and 3D). For shallow LCPS, ORA revealed the enrichment in pathways such as gamma-carboxylation processes and extracellular matrix (ECM) organization. These pathways reflect acute COVID-related processes, as ECM degradation and inflammation-driven protease activity are characteristic of the disease [24,25]. Gamma-carboxylation, indirectly linked to coagulation abnormalities and fibrinogen dysregulation, is also implicated in COVID-19 pathology. In contrast, deep LCPs were enriched for pathways associated with complement regulation, thromboinflammatory mechanisms, and immune integrations (Figure 3D), highlighting their direct relevance to Long COVID mechanisms [23,26]. These findings demonstrate that shallow LCPs primarily reflect the general processes of COVID-19, while deep LCPs capture pathways that are mechanistically essential for the long-term effects of Long COVID. This is plausible, as Long COVID is less related to acute interactions with the initial SARS-CoV-2 virus. This underscores the necessity of analyzing the network at all depths to gain a comprehensive understanding of the disease progression toward Long COVID.

In order to evaluate the ability of QW and RWR to identify LCPs across different network depths, we compared the spatial distributions of LCPs with those of the key proteins predicted by QW and RWR within the SIP network (Figure 3E). Visualized networks revealed that the spatial distribution of QW key proteins closely resembled that of LCPs, whereas RWR key proteins exhibited a markedly different pattern. LCPs were predominantly located in deep nodes beyond the DIPs, a spatial distribution that was similarly reflected in QW key proteins. In contrast, RWR key proteins were concentrated near the DIPs in shallow nodes, with no representation at distance of 4, further highlighting the limitations of RWR in exploring deep regions (Figure 3E).

This disparity was further highlighted by the predicted probability distributions for all LCPs (Figure 3F). While the probability distribution for shallow LCPs showed no significant difference between the two methods, QW assigned significantly higher probabilities to deep LCPs compared to RWR. Thus, QW demonstrated greater effectiveness in identifying both shallow and deep LCPs, likely due to its enhanced capability to explore deeper regions of the network. Considering our finding that deep LCPs-enriched pathways are directly implicated in the mechanisms of Long COVID, these results suggest that QW is more likely to identify key LCPs that are strongly associated with the pathophysiology of Long COVID.

**QW outperforms RWR in uncovering Long COVID mechanisms and enriched pathways**

To assess the performance of QW and RWR in identifying LCPs, we compared the overlap between the 947 key proteins discovered by both methods and the LCPs located in shallow and deep nodes (Figure 4A). In shallow nodes, QW identified 24 LCPs, while RWR identified 32, with no significant difference observed between the methods (Supplementary Table 1). The overlap between shallow LCPs and key proteins showed a similar pattern across QW and RWR, as indicated by the intersections in the Venn diagram in Figure 4A. ORA of QW-predicted shallow LCPs revealed associations with Wnt/β-catenin signaling, apoptotic cleavage, and hormone metabolism pathways (Supplementary Figure B). Similarly, RWR-predicted pathways included Wnt/β-catenin signaling and uniquely highlighted transcriptional regulation, cell surface interactions, and pathogen recognition pathways (Supplementary Figure C). These pathways, reflecting general cellular processes and post-infection remodeling, appear to have less relevance to Long COVID mechanisms.

In contrast, QW outperformed RWR in deep nodes by uniquely identifying 27 LCPs that were absent from RWR predictions. These 27 proteins are represented in the unique intersection of the Venn diagram for deep LCPs (Figure 4A). ORA of these deep LCPs revealed enrichment in Long COVID-related pathways, including persistent dysregulation of the complement cascade, which contributes to thromboinflammation, chronic inflammation, and vascular injury [23,26,27] (Figure 4B). Additional enriched pathways included coagulation and fibrin clot formation, which are closely associated with hallmark Long COVID symptoms such as fatigue and post-exertional malaise [23,26]. To further compare QW and RWR at the pathway level, we assessed their ability to identify pathways strongly enriched in LCPs (Supplementary Figure D, E). ORA was performed on shallow and deep LCPs, QW key proteins, and RWR key proteins, and their overlaps were analyzed. Supplementary Figure D highlights the enriched pathways identified by both QW and RWR for shallow LCPs, revealing that QW captures pathways such as extracellular matrix organization, blood coagulation, and protein modification, while RWR focuses on signal transduction and cellular regulation. Similarly, Supplementary Figure E illustrates the pathways enriched for deep LCPs, where QW uniquely emphasizes thromboinflammatory responses, immune regulation signaling, and glycosylation-related processes, compared to the narrower focus of RWR. These results provide a clear comparison of how each method performs across shallow and deep regions of the network.

In Figure 4C, QW-enriched pathways include neuronal inflammation, metabolic inflammation, and glycosylation disease pathways, which are strongly linked to Long COVID manifestations. In contrast, Figure 4D shows that RWR-enriched pathways predominantly highlight processes like nucleolar function, viral replication, and nonsense-mediated mRNA decay (NMD). The findings demonstrated that, for the top 10 enriched pathways of both shallow and deep LCPs, QW identified more pathways with direct relevance to Long COVID mechanisms than RWR. An evaluation of QW and RWR's capability to investigate Long COVID mechanisms was conducted by focusing on the top 10 statistically significant pathways from the ORA results of the 947 proteins predicted by both approaches. Pathways identified by QW are visualized in Figure 4C, with key links to neuronal inflammation [28-30], metabolic inflammation [31], glycosylation disease [32], and FGF signaling dysregulation [33]. In contrast, to the best of our knowledge, none of the top 10 enriched pathways identified by RWR (Figure 4D) were found to have a direct association with Long COVID symptoms.

To investigate the pathways enriched by QW and RWR in the SIP network, we employed a circos plot (Figure 4E and 4F). This visualization effectively highlights the distribution of enriched pathways across shallow nodes, which are located closer to the SARS-CoV-2, and deep nodes, representing downstream and systemic effects of the viral perturbation. The circos plot divides the SIP network into distinct pathway categories enriched by each method. Pathways enriched by QW are represented in red and orange hues, while those enriched by RWR are shown in blue and green hues. Radial lines within the plot indicate the connections between nodes (proteins) and their enriched pathways, demonstrating how these pathways span across shallow and deep regions of the network. The outermost ring categorizes nodes based on their distance to the SARS-CoV-2 node: shallow nodes reflect direct, immediate effects of the virus, while deep nodes capture downstream, systemic effects.

Figure 4E illustrates that QW-enriched pathways span both shallow and deep nodes, with red and orange hues representing key processes like glycosylation, synapses, and neuronal inflammation. These pathways are associated with immune regulation, chronic inflammation, and neurological symptoms, which are hallmark features of Long COVID. The distribution of QW-enriched pathways across both shallow and deep regions suggests its sensitivity to both localized effects near the viral signal and cascading effects further downstream. In addition, the QW circos plot (Figure 4E) exhibited extensive connectivity patterns, particularly in pathways associated with neuronal inflammation (orange and yellow band) [28-30], glycosylation diseases (red band) [34], and glycosaminoglycan metabolism (pink band) [35]. These pathways showed strong associations with autoimmune responses characteristic of Long COVID phenotypes (Figure 4C) [36]. The enriched pathways reinforce the relevance of QW in capturing key processes linked to immune dysregulation and persistent inflammation that underpin Long COVID [37].

In contrast, RWR circus plot (Figure 4F) predominantly highlights shallow nodes enriched in pathways related to viral replication, nonsense-mediated mRNA decay (NMD), and nucleolar stress. NMD is a highly conserved cellular surveillance pathway that plays a central role in mRNA quality control and gene expression regulation. During COVID-19, NMD may become disrupted, impairing mRNA surveillance and contributing to immune dysregulation and cellular stress [38]. This disruption has been linked to prolonged immune activation and inflammation, processes potentially underlying Long COVID [34]. Pathways related to nucleolar function also emerged, reflecting the impact of COVID-19 on ribosomal RNA (rRNA) synthesis [39]. Viral proteins such as NSP1 and NSP5 localize to the nucleolus, inducing nucleolar stress by disrupting rRNA transcription and ribosome biogenesis [40]. This disruption not only impairs protein synthesis but also enables the virus to hijack nucleolar functions to enhance replication. The resulting nucleolar stress and its downstream effects on cellular homeostasis and immune responses are likely contributors to the persistent symptoms seen in Long COVID [41].

While RWR-enriched pathways highlight key processes associated with the direct effects of viral infection and host responses, their predominance in shallower regions suggests that RWR primarily captures immediate, localized interactions rather than systemic, long-range perturbations. This complements the broader, deep-region focus observed with QW, offering a complementary perspective on COVID-19's impact on host biology.

**QW links mitochondrial dysfunction to Long COVID mechanisms and identifies novel therapeutic targets**

To assess the ability of the QW approach to propose specific Long COVID mechanisms from the SIP network and further identify novel therapeutic biomarkers, we examined the top 10 proteins predicted by QW (Figure 5A and Supplementary Table 2). Interestingly, the probability values for all proteins, except for TP53, were significantly lower in RWR compared to QW, emphasizing that these prioritized proteins were uniquely identified through the QW approach. Of particular interest was the inclusion of CISD1, CISD2 and CISD3 proteins from the CDGSH iron-sulfur binding domain-containing family in the list. Figure 5A highlights the significant difference in normalized probabilities between QW and RWR, with QW strongly prioritizing CISD family proteins. The CDGSH domain, which is a 2Fe-2S iron-sulfur binding motif, is known to participate in various reactions that facilitate the transfer of electrons or iron-sulfur clusters to other receptor proteins [42]. Notably, CISD1, CISD2 and CISD3 are particularly involved in regulating reactive oxygen species (ROS) metabolism [43]. The specific association between ROS metabolism and iron-sulfur clusters with COVID-19 and Long COVID is of significant interest, as multiple reports have highlighted these connections [44-46].

We focused on this protein family and extracted a subnetwork of proteins directly connected to the CISD1, CISD2 and CISD3 proteins in the SIP network (Figure 5B). The subnetwork was highly interconnected, with most of the proteins localized in the mitochondrion membrane and endoplasmic reticulum (ER) membrane, suggesting that the interactions and functions of the proteins are closely related to mitochondria-associated membranes (MAMs). Figure 5B provides a schematic view of the intracellular locations of these proteins, highlighting the dense connectivity within the MAMs and their involvement in calcium ($Ca^{2+}$) regulation. WFS1, CISD2, BCL2, and VDAC1 play critical roles in regulating $Ca^{2+}$ concentrations between the ER and mitochondria within the MAMs [47]. CANX (calnexin) also stimulated the ATPase activity of SERCA at the MAMs, which enabled the control of ER $Ca^{2+}$ availability for mitochondria [48]. Dysregulated mitochondrial Ca²⁺ levels can lead to structural collapse of mitochondria or trigger mitophagy [49,50].

BECN1 (Beclin1) plays a pivotal role in mitophagy by localizing to MAMs, where it ensures autophagosome formation [51]. PRKN (PARK2) is essential for mitophagy through its ability to promote mitochondrial ubiquitylation and recruit ubiquitin-binding mitophagy receptors [52]. Additionally, a study has reported that BECN1 facilitates PRKN translocation to mitochondria, thereby contributing to the regulation of mitophagy [53]. To further explore the overall biological processes associated with these proteins, we performed GO based ORA, revealing significant enrichment in autophagy, mitophagy, ROS metabolism, and endoplasmic reticulum stress response (Figure 5C and Supplementary Table 3). Thus, an intensive literature review of each individual protein, coupled with computational enrichment analyses of all identified proteins, strongly supports the association between the QW subnetwork and mitochondrial dysfunction and mitophagy. Indeed, mitochondrial dysfunction and mitophagy have been identified as a key factor in the pathogenesis of Long COVID [55-56].

Finally, we investigated whether QW could specifically identify drug targets for Long COVID. Within the subnetwork shown in Figure 5B, VDAC1 emerged as a potential target due to its critical role in mitochondrial Ca²⁺ transport and its significant reduction in Long COVID patients [57,58]. The key function of the subnetwork we identified was the regulation of Ca²⁺ at the MAMs, with particular emphasis on VDAC1, a voltage-dependent anion channel protein that directly participates in mitochondrial Ca²⁺ transport. To evaluate VDAC1 as a potential therapeutic target, we investigated compounds targeting VDAC1 using the DrugBank database. Among the four identified compounds, cannabidiol is the only FDA-approved drug. Notably, the potential of cannabidiol as a therapeutic agent for Long COVID has been explored, leveraging its neuropsychiatric properties and anti-inflammatory effects [59-61].

## Discussion

Although from a public health perspective, COVID-19 has transitioned from a pandemic to an endemic disease, many patients still suffer from Long COVID, and clear solutions remain unavailable. In this study, we demonstrate that QW offers a deeper exploration of the global structure of large-scale SIP networks compared to RWR, enabling more precise identification of proteins and pathways implicated in Long COVID pathophysiology. Additionally, QW-based analyses highlight its potential for uncovering crucial disease pathway (mitochondrial dysfunction) and novel therapeutic targets, advancing our understanding of the Long COVID mechanisms. These findings underscore the potential of QW as a robust computational approach for investigating complex biological systems, positioning it as a promising tool for the study of multisystem disorders such as Long COVID.

One of the major limitations of the conventional network analysis method RWR is its inability to effectively explore deeper regions of biological networks. This constraint was evident in the present study, where RWR primarily identified proteins located near the SARS-CoV-2 node, failing to detect critical interactions in more distal regions of the network. In contrast, QW, by leveraging quantum interference, demonstrated superior network traversal capabilities, assigning higher probabilities to nodes in deeper network layers. This ability to probe distal and systemic regions of the network offers a unique advantage in exploring conditions like Long COVID. Since Long COVID occurs later in the disease process compared to the initial infection, key biological processes are likely to be distributed across more diverse and distal regions of the network.

Our analysis corroborates previous research on QW's utility in network analysis, demonstrating its superior capability to prioritize LCPs over RWR. Specifically, LCPs were observed to be distributed not only in shallow regions but also in deeper regions of the network. Notably, the distribution of key proteins predicted by QW closely mirrored the distribution of LCPs, showing a significantly higher concordance compared to RWR. Consequently, key proteins identified by QW proved to be more effective in broadly and accurately recapitulating the mechanisms suggested by LCPs in Long COVID pathophysiology.

The comparative pathway enrichment analysis between QW and RWR further underscores the unique advantages of QW in elucidating Long COVID mechanisms. QW identified pathways related to thromboinflammatory responses and neuronal inflammation, which are directly linked to Long COVID symptoms such as fatigue, cognitive dysfunction, and systemic inflammation. Moreover, QW uniquely highlighted mitochondrial dysfunction as a key process, further connecting Long COVID to systemic dysregulation in energy metabolism. In contrast, RWR predominantly enriched pathways associated with acute COVID-19 dynamics, highlighting its limited utility for capturing the chronic and systemic nature of Long COVID.

One particularly interesting aspect of our findings is the close relationship between the proteins within the subnetwork we identified and Parkinson's disease. Notably, mutations in the PRKN gene are a well-established cause of Parkinson's disease [62]. Since the onset of the COVID-19 pandemic, potential links between COVID-19 and Parkinson's disease have garnered attention [63]. Recent studies suggest that COVID-19 infection could increase the risk of developing Parkinson's disease or parkinsonism, potentially due to shared mechanisms involving mitochondrial dysfunction, immune dysregulation, and chronic inflammation [64].

Furthermore, the suggestion of VDAC1 as a potential therapeutic target for Long COVID is promising, given its critical role in mitochondrial $Ca^{2+}$ regulation. Previous studies have identified VDAC1 as a biomarker for Long COVID due to its altered expression in affected patients. Our findings expand on these observations by linking VDAC1 to mitochondrial dysfunction and systemic inflammation, providing further evidence of its therapeutic relevance. In addition, VDAC1 has been extensively studied as a potential therapeutic target for inflammation-related diseases and neurodegenerative disorders, such as Alzheimer's disease. Considering the symptomatic similarities between Long COVID and these conditions, our findings indicate that our discovery may have significant clinical relevance [65-67].

While this study demonstrates the utility of QW in Long COVID research, several limitations warrant consideration. The constructed protein interaction network relies on available databases and may not fully capture the dynamic and context-specific interactions underlying Long COVID. Incorporating longitudinal and multi-omics data could further enhance the accuracy and relevance of the network. Additionally, the performance of QW in biological networks is influenced by the choice of operator and parameters, which may impact its ability to capture subtle interactions. Future studies should focus on integrating QW with experimental approaches, such as proteomics and functional assays, to validate and extend these findings.

This study underscores the transformative potential of QW in advancing our understanding of Long COVID. By enabling deeper exploration of biological networks, QW has identified novel proteins and pathways that are critically involved in Long COVID pathophysiology. In particular, its ability to prioritize mitochondrial dysfunction and systemic processes further establishes its value in studying multisystem disorders or other post-acute sequelae, such as those following influenza, Ebola, or dengue [68-70]. These findings not only enhance our understanding of this complex condition but also pave the way for the development of targeted therapies. As computational biology continues to evolve, approaches like QW hold promise for addressing the challenges posed by multifaceted diseases, ultimately contributing to more effective and personalized healthcare solutions.

## Materials and Method

### Directly interacting proteins, differentially expressed proteins

A total of 332 high-confidence SARS-CoV-2–human interactions were obtained from Gordon *et al.* (table S3 from https://doi.org/10.1038/s41586-020-2286-9). A total of 332 high-confidence virus-host interactions were used as DIPs. Data of differentially expressed SomaScan measurements in 6-month Long Covid patients (compared to recovered COVID-19 patients) at acute COVID-19 (defined as DEPs) were obtained from Cervia-Hasler et al. A total of 1335 DEPs were extracted by combining Data S1 and S3 (from DOI: 10.1126/science.adg7942). The proteins that were significantly up- or down- regulated (two-tailed t tests, $P < 0.05$, $|\log_2 FC| > 0$) were selected.

### Long COVID network construction

Long COVID network was constructed of all shortest paths between DIPs and DEPs in a human protein-protein interaction network from STRING database [71]. Only interactions with a confidence score of greater more than medium (0.4) were used. All shortest paths between DIPs and DEPs were found using the Python package NetworkX [72].

### Long Covid Proteins

Data of differentially expressed SomaScan measurements in 6-month Long Covid patients (compared to recovered COVID-19 patients) at 6-month follow-up (defined as LCPs) were obtained from Cervia-Hasler et al. A total of 514 LCPs were extracted by combining Data S2 and S4 (from DOI: 10.1126/science.adg7942). The proteins that were significantly up- or down- regulated (two-tailed t tests, $P < 0.05$, $|\log_2 FC| > 0$) were selected.

### Network analysis

QW and RWR were employed to prioritize Long COVID related proteins in the SIP network. The objective of this analysis was to identify key proteins in the pathway starting from SARS-CoV-2 nodes to DEPs. Therefore, the SARS-CoV-2 node was used as the starting node for both QW and RWR. RWR was implemented using the Pagerank algorithm from the Python package NetworkX [72], with the default parameters provided by the package (maximum iteration number: 100, restart probability: 0.15, error tolerance: $1.0 \times 10^{-6}$). QW was implemented using the Coined algorithm from the Python package Hyperwalk, with the evolution operator constructed using the Grover coin and the flip-flop shift operator [73]. Permutation tests were performed 1,000 times to identify significant proteins for each of the network centrality algorithms. For each permutation test, a random network that has the same degree distribution as the SIP network was generated. If a protein has less than permutation p-value 0.01 for each of the network centrality algorithms, we considered it a key protein. Distance analysis was performed using the Shortest Paths function of NetworkX. The single farthest node with a distance of 6 from the SARS-CoV-2 node was excluded from the histogram visualization.

### Calculation of EDP and EDAP

To calculate the change in probability distributions at each iteration from the QW and RWR analysis results, the Euclidean distance of probability distribution was used.

Let $G = (V, E)$ be the analyzed graph with $|V| = n$. And let $v$ be a node of graph $G$, then EDP at step t is defined as

$$EDP_t = \sqrt{\sum_{i=1}^{n}(P_t(v_i) - P_{t-1}(v_i))^2}$$

Where $P_t(v_i)$ is probability on the node $v_i$ at step $t$. To address the oscillatory behavior of QW, which does not converge, the averaged probability and EDAP of node v is at step t is defined as

$$\bar{P}_t(v) = \frac{1}{t+1} \sum_{k=0}^{t} P_k(v)$$

$$EDAP_t = \sqrt{\sum_{i=1}^{n} (\bar{P}_t(v_i) - \bar{P}_{t-1}(v_i))^2}$$

where $n$ is the total number of nodes in the network. Finally, the average mixing time is defined as

$$M_\epsilon = \min \{T | \forall t \geq T \rightarrow EDAP_t(V) \leq \epsilon\}$$

where $\epsilon$ is arbitrary small value. We used a value of $1.0 \times 10^{-5}$ in this analysis, based on our observation that the EDAP converged near this value. The calculated averaged mixing time for QW was 1,212 iterations.

**Over-Representation Analysis (ORA)**

Over-representation analysis (ORA) performed using the R package clusterProfiler (v4.14.4) [74]. Two distinct ORAs conducted using Gene Ontology (GO) and REACTOME gene set databases. REACTOME-based ORA of LCPs applied an adjusted p-value threshold of < 0.01 to identify significantly enriched pathways. REACTOME-based ORA of 27 overlapping proteins between QW key proteins and deep LCPs applied an adjusted p-value threshold of < 0.01. REACTOME-based ORA of QW and RWR key proteins extracted the top 10 enriched pathways based on adjusted p-values. GO-based ORA of the CISD subnetwork extracted the top 20 enriched pathways based on adjusted p-values.

# Acknowledgements

**Funding**

J.P. is funded by Helmsley Charitable Trust, the Leona M and Harry B Helmsley Chartiable Trust grant (grant number G118500). W.H. and M.M. are funded by CardiaTec Biosciences. S.L. PhD studentship is jointly funded by Cambridge Stem Cell Institute and Milner Therapeutics Institute. H.L. PhD studentship is funded by Milner Therapeutics Institute. N.H. is funded by LifeArc.

**Author contributions**

Conceptualization: J.P., W.H., N.H.
Methodology: J.P., W.H., M.M.
Investigation: J.P., W.H., S.L., H.L.
Visualization: J.P., W.H., S.L., H.L., N.H.
Supervision: N.H.,
Funding: M.Z., N.H.
Writing—original draft: J.P., W.H., S.L., H.L.
Writing—review & editing: J.P., W.H., N.H.

**Competing interests**

N.H. is a cofounder of KURE.ai and CardiaTec Biosciences. W.H and M.M. are employees of CardiaTec Biosciences. All other authors declare that they have no competing interests.

**Data and materials availability**

All data needed to evaluate the conclusions in the paper are present in the paper and/or the Supplementary Materials. The code and input data that were used for this study are available at TBA. Additional data related to this paper may be requested from the authors.

# Figures

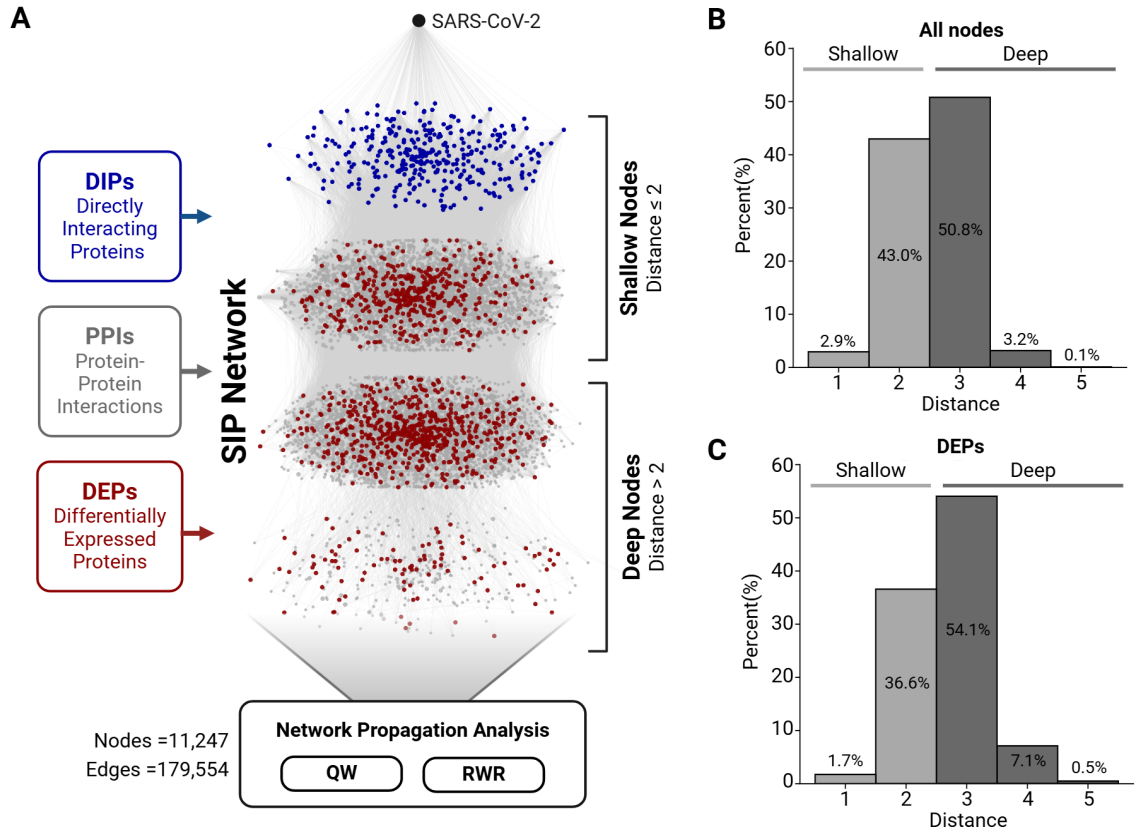

**Figure 1. The SIP network distances show that deep exploration is needed to analyze paths to DEPs. (A)** Schematic diagram of the SIP network construction process. The central network represents the constructed SIP network, with three clusters of nodes at the top corresponding to shallow nodes (distances 1, 2, and 3 from the SARS-CoV-2 node) and one cluster at the bottom representing deep nodes (distances ≥ 4). **(B)** Histogram showing the distance distribution of all nodes in the SIP network according to their distance from the SARS-CoV-2 node. **(C)** Histogram illustrating the distance distribution of DEPs in the SIP network from the SARS-CoV-2 node, highlighting their enrichment in deep regions.

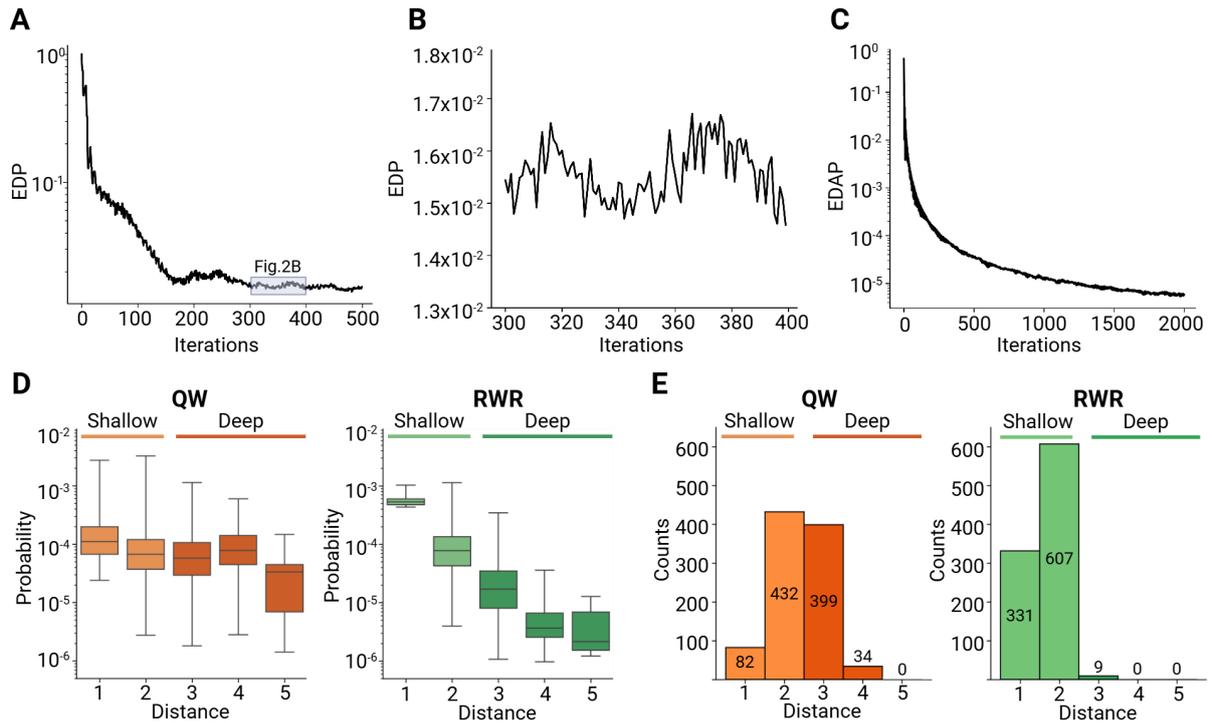

**Figure 2. Analysis of QW and RWR on the SIP network reveals a fundamental difference in their exploration depth**. **(A)** EDP at each iteration in QW, with the y-axis presented on a logarithmic scale to capture the wide dynamic range of changes. **(B)** EDP between the 300th and 400th iterations in QW, highlighting the oscillatory behavior that indicates non-convergence. **(C)** EDAP at each iteration in the QW, with the y-axis presented on a logarithmic scale. **(D)** Probability distribution of all nodes based on their distances from the SARS-CoV-2 node. The results of the QW are shown on the left, while those of the RWR are displayed on the right. **(E)** Histogram of top 947 nodes based on their distances from the SARS-CoV-2 node. These findings demonstrate QW's ability to explore deeper regions of the network compared to RWR.

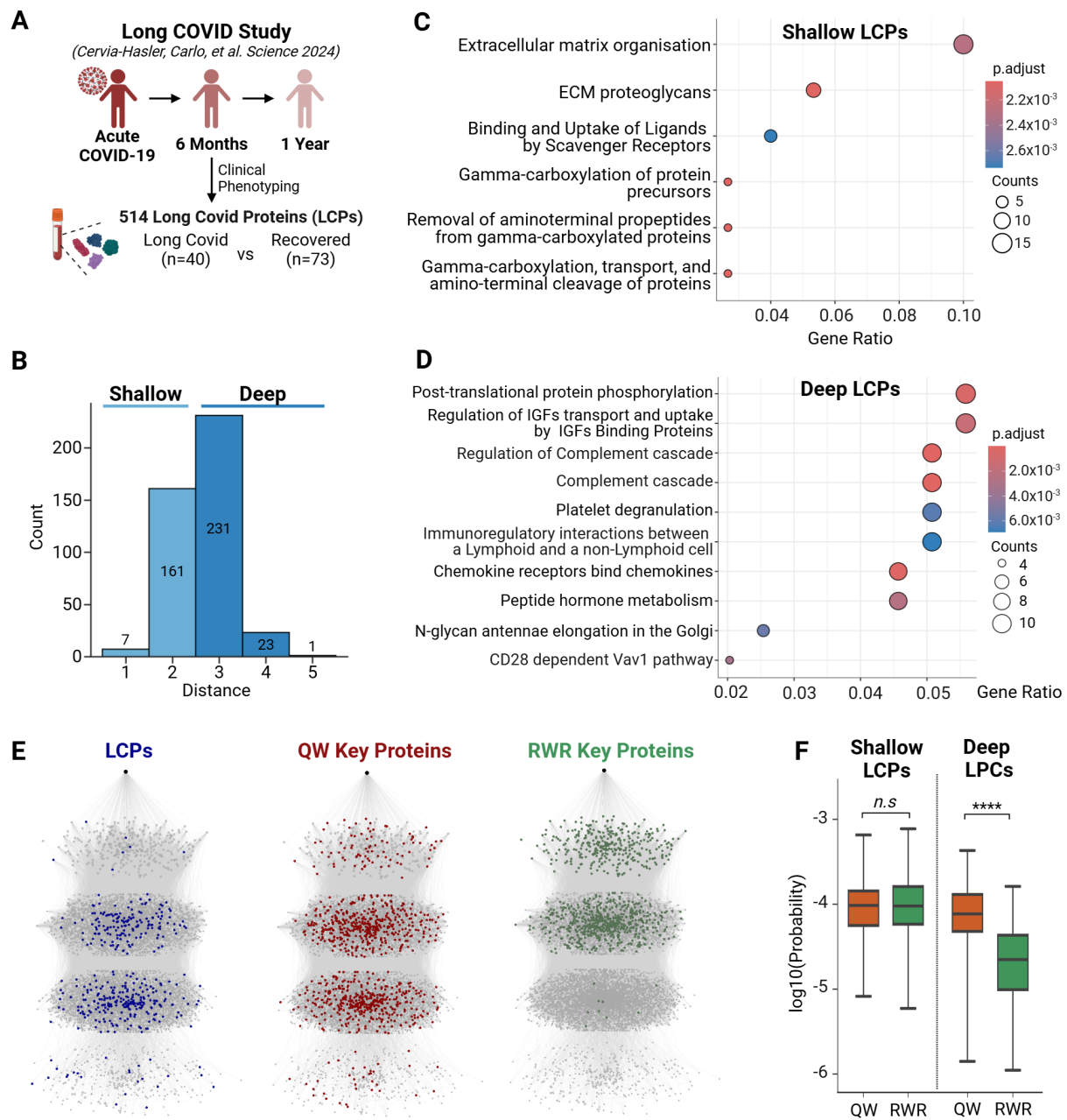

**Figure 3. QW-identified key proteins closely reflect the spatial distribution of LCPs in the SIP network, outperforming RWR in identifying deep LCPs. (A)** Schematic representation of a 12-month longitudinal proteomics study by Cervia-Hasler et al., *Science,* 2024. Blood samples were collected at 6 months post infection from Long COVID patient (n=40) and recovered COVID-19 patients (n=73), leading to the identification of 514 differentially expressed proteins LCPs by comparing two groups. **(B)** Spatial distribution of LCPs within the Long COVID network. Of the 423 mapped LCPs, 60% were positioned in deep nodes (distance >2) compared to 40% in shallow nodes (distance ≤2)., reflecting their relevance to both acute and long-term disease processes. **(C)** REACTOME-based ORA results for shallow LCPs, highlighting pathways such as ECM organization and gamma-carboxylation processes, linked to acute COVID-related processes. **(D)** REACTOME-based ORA results for deep LCPs, highlighting pathways related to complement regulation and thromboinflammatory mechanisms, critical to Long COVID pathophysiology. **(E)** Visualisation of the Long COVID network showing the spatial distribution of LCPs, QW key proteins, and RWR key proteins. Node are grouped into clusters based on their distance from the starting node (SARS-CoV-2): 1, 2, 3, and 4+. **(F)** Comparison of node probability for LCPs between QW and RWR across shallow and deep nodes. While shallow node probabilities showed no significant difference (n.s.), QW assigned significantly higher probabilities to deep LCPs than RWR (****).

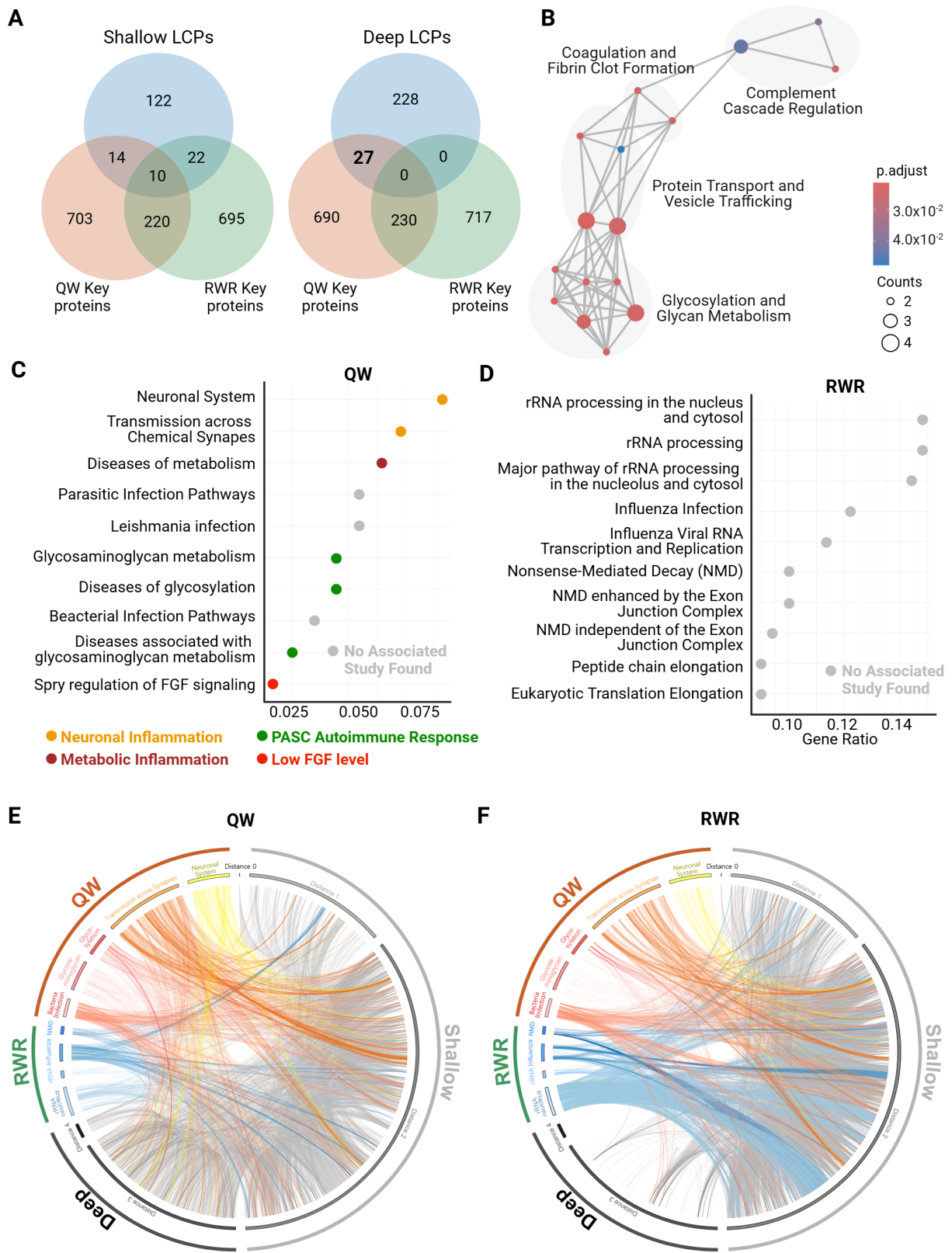

**Figure 4. QW and RWR reveal distinct mechanisms underlying Long COVID symptoms, with QW demonstrating superior performance.** (**A**) Venn diagram of overlapping proteins between shallow and deep LCPs, QW key proteins, and RWR key proteins. QW uniquely identified 27 deep nodes overlapping with LCPs, which were not detected by RWR. (**B**) REACTOME-based ORA results for the 27 deep LCPs uniquely identified by QW. Pathways sharing 20% of proteins are connected by edges. (**C**) REACTOME-based ORA results for the

947 key proteins of QW. Pathways are coloured based on their functional similarity, with many linked to Long COVID symptoms such as inflammation and metabolic dysregulation. **(D)** REACTOME-based ORA results for the 947 key proteins of RWR, emphasizing processes like nucleolar stress, NMD, and viral replication. **(E)** A circos plot depicting the interactions between key proteins identified by QW and the QW-enriched pathways (red/orange) as well as RWR-enriched pathways (blue/green). The key proteins are subdivided based on their distance from SARS-CoV-2. **(F)** A circos plot depicting the interactions between key proteins identified by RWR and the QW-enriched and RWR-enriched pathways. The key proteins are subdivided based on their distance from SARS-CoV-2.

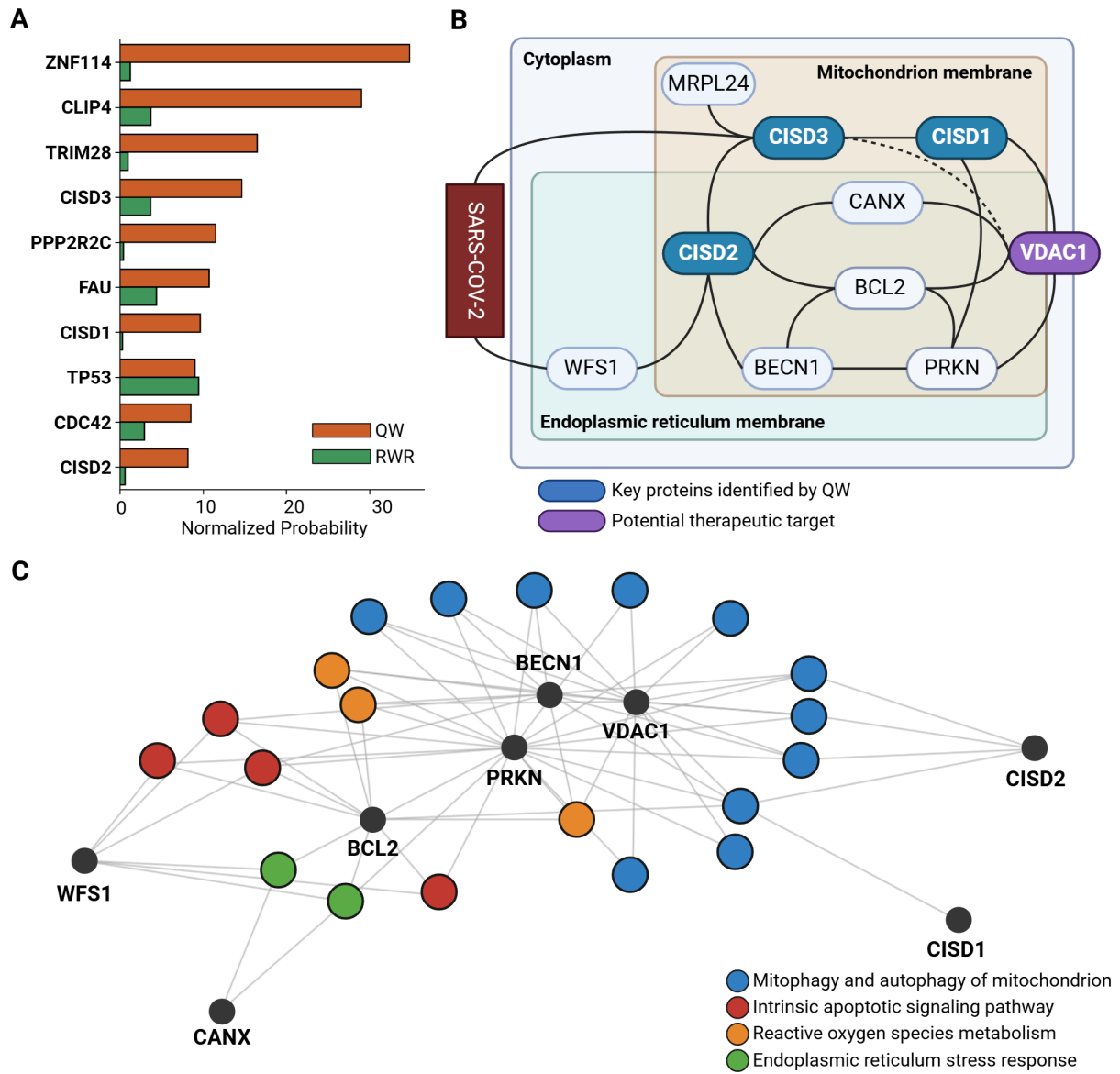

**Figure 5. The top 10 QW-identified proteins reveal Long COVID mechanisms and therapeutic targets. (A)** Z-normalized probability of the top 10 proteins predicted by QW in the SIP network. The QW results are displayed in an orange, and the RWR results are displayed in a green for comparison, showing that QW prioritizes most proteins significantly more than RWR. **(B)** Visualization of a sub-network comprising proteins directly connected to CISD1, CISD2, and CISD3 in the SIP network. The rectangular areas schematically represent the intracellular locations of proteins: the blue area denotes the cytoplasm, the green area denotes the endoplasmic reticulum membrane, and the brown area denotes the mitochondrial membrane. Solid lines represent connections within the SIP network, while dashed lines denote additional direct associations confirmed through literature evidence. **(C)** The network visualization illustrates the linkages between proteins in the subnetwork and their associated biological processes, based on the top 20 results from Gene ontology (GO) based ORA of these proteins. Biological processes associated with the proteins are clustered and displayed in the same color to represent functional similarity. Processes include autophagy, mitophagy, ROS metabolism, and ER stress response, emphasizing their relevance to Long COVID mechanisms.

# Supplementary Materials

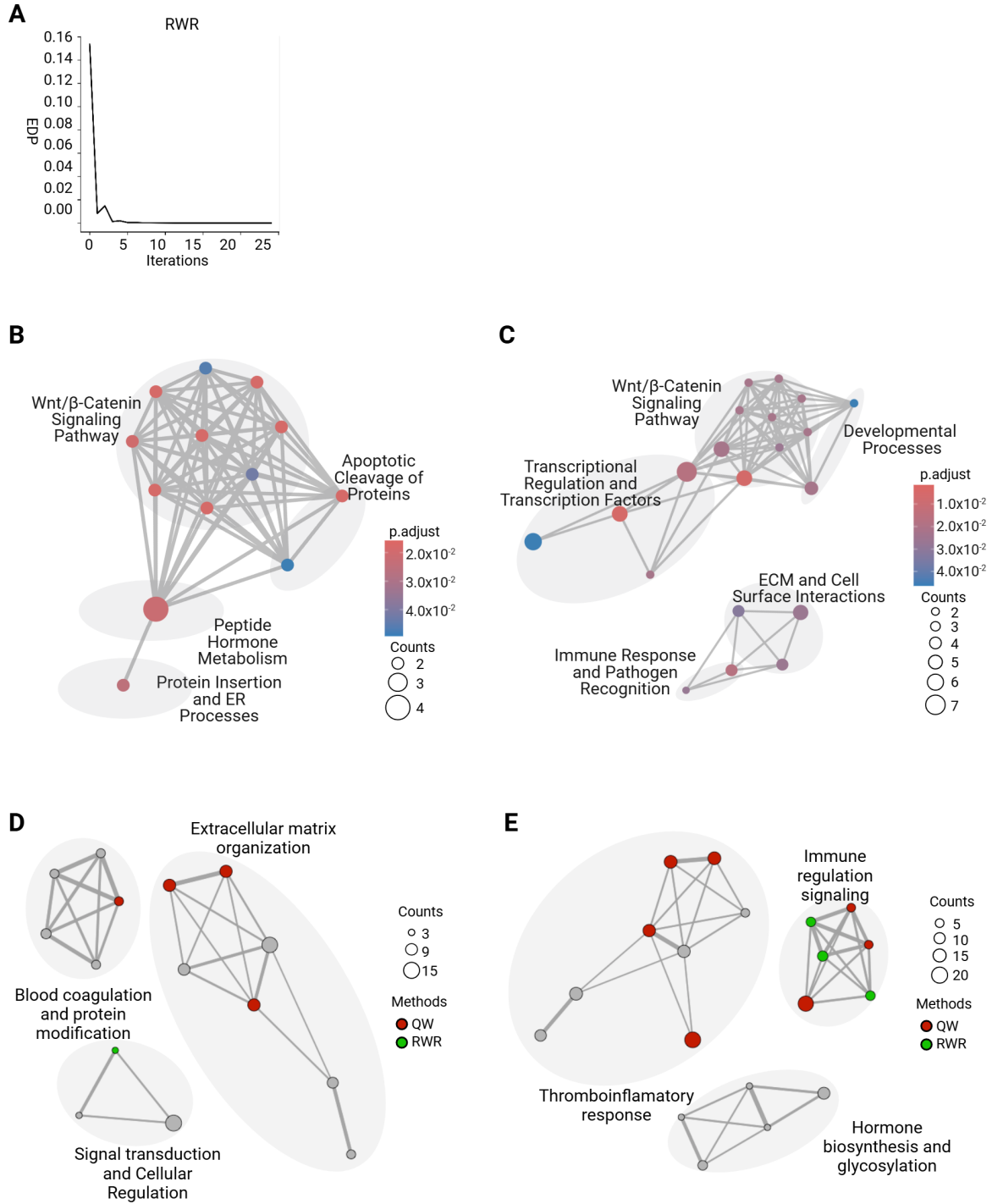

**Supplementary Figure.**

**(A)** EDP at each iteration in RWR. **(B)** REACTOME-based ORA results for the 24 shallow LCPs identified by QW. **(C)** REACTOME-based ORA results for the 32 shallow LCPs identified by RWR. **(D)** REACTOME-based ORA results for all shallow LCPs. The three largest pathway clusters are represented. The enriched pathways identified from the ORA analysis of key proteins predicted by QW and RWR were represented in red and green, respectively. **(E)** REACTOME-based ORA results for all deep LCPs. The three largest pathway clusters are represented. The enriched pathways identified from the ORA analysis of key proteins predicted by QW and RWR were represented in red and green, respectively.

**Supplementary Table 1.** Intersecting protein lists between QW & RWR key proteins, and LCPs

| QW Key Proteins & Shallow LCP | RWR Key Proteins & Shallow LCP | QW Key Proteins & Deep LCP |
|---|---|---|
| CTNNB1 | EGFR | AGT |
| APP | CTNNB1 | FOLH1 |
| EGFR | ACADM | PANK1 |
| NPY | RALA | KNG1 |
| ACADM | APP | TG |
| FXN | PLAT | NUDT9 |
| GABARAP | NEU1 | UNC5D |
| GABARAPL1 | GDF15 | SFTPD |
| STX1A | LMAN2 | NUDT15 |
| F7 | QSOX2 | B4GALT1 |
| RALA | CREBBP | RGMA |
| GDF15 | POLR1C | LMAN2L |
| NOVA1 | GSK3B | B4GALT2 |
| GSK3B | HSP90B1 | SCG5 |
| DSG1 | GABARAP | FCRL5 |
| ITSN1 | RAD51 | C3 |
| ATP1B2 | SNRPG | AKR1A1 |
| LUM | CXCL8 | PDXK |
| GAL3ST1 | WDR5 | CTSV |
| MFAP5 | GABARAPL1 | GRID2 |
| VWF | NPM1 | PTPRD |
| FMR1 | COL3A1 | GRB7 |
| GHRL | STX6 | CPN1 |
| ESAM | RRM2 | CST6 |
| | FMR1 | IGF2R |
| | COX5A | F8 |
| | CHEK1 | B3GAT3 |
| | ICAM1 | |
| | AURKB | |
| | KIT | |
| | KDR | |
| | SPARC | |

**Supplementary Table 2.** Top 10 proteins of QW result

| Name | QW rank | RWR rank | Cell location | QW probability (Z-normalized) | RWR probability (Z-normalized) |
|---|---|---|---|---|---|
| ZNF114 | 2 | 838 | Nucleus | 0.003263 | 0.000213 |
| CLIP4 | 3 | 255 | Nucleus, Cell cortex | 0.002736 | 0.000491 |
| TRIM28 | 4 | 1064 | Nucleus | 0.001594 | 0.000183 |
| CISD3 | 5 | 261 | Mitochondrion | 0.001426 | 0.000487 |
| PPP2R2C | 6 | 1781 | Cytosol | 0.001140 | 0.000123 |
| FAU | 7 | 142 | Nucleus, Cytoplasm | 0.001065 | 0.000566 |
| CISD1 | 8 | 2083 | Mitochondrion outer membrane | 0.000969 | 0.000108 |
| TP53 | 9 | 2 | Nucleus, Cytoplasm, ER, Mitochondrion matrix | 0.000911 | 0.001136 |
| CDC42 | 10 | 388 | Cell membrane, Cytoplasm | 0.000865 | 0.000405 |
| CISD2 | 11 | 1440 | Mitochondrion outer membrane, ER | 0.000832 | 0.000145 |

**Supplementary Table 3. Top 20 GO ORA result of the CISDs related subnetwork**

| Ranking | GO id | Term description | Adjusted p-value |
|---|---|---|---|
| 1 | GO:2000378 | negative regulation of reactive oxygen species metabolic process | $3.48 \times 10^{-6}$ |
| 2 | GO:0010506 | regulation of autophagy | $3.48 \times 10^{-6}$ |
| 3 | GO:0000422 | autophagy of mitochondrion | $2.36 \times 10^{-5}$ |
| 4 | GO:0061726 | mitochondrion disassembly | $2.36 \times 10^{-5}$ |
| 5 | GO:0098780 | response to mitochondrial depolarisation | $2.36 \times 10^{-5}$ |
| 6 | GO:1903008 | organelle disassembly | $8.50 \times 10^{-5}$ |
| 7 | GO:2000377 | regulation of reactive oxygen species metabolic process | $8.50 \times 10^{-5}$ |
| 8 | GO:0000423 | mitophagy | $1.01 \times 10^{-4}$ |
| 9 | GO:2001242 | regulation of intrinsic apoptotic signaling pathway | $1.84 \times 10^{-4}$ |
| 10 | GO:0072593 | reactive oxygen species metabolic process | $3.48 \times 10^{-4}$ |
| 11 | GO:0070059 | intrinsic apoptotic signaling pathway in response to endoplasmic reticulum stress | $3.48 \times 10^{-4}$ |
| 12 | GO:0034976 | response to endoplasmic reticulum stress | $5.25 \times 10^{-4}$ |
| 13 | GO:0098779 | positive regulation of mitophagy in response to mitochondrial depolarization | $7.25 \times 10^{-4}$ |
| 14 | GO:0061912 | selective autophagy | $8.60 \times 10^{-4}$ |
| 15 | GO:0097193 | intrinsic apoptotic signaling pathway | $8.60 \times 10^{-4}$ |
| 16 | GO:1904925 | positive regulation of autophagy of mitochondrion in response to mitochondrial depolarization | $1.10 \times 10^{-3}$ |
| 17 | GO:0036503 | ERAD pathway | $1.10 \times 10^{-3}$ |
| 18 | GO:2001243 | negative regulation of intrinsic apoptotic signaling pathway | $1.10 \times 10^{-3}$ |
| 19 | GO:1903599 | positive regulation of autophagy of mitochondrion | $1.10 \times 10^{-3}$ |
| 20 | GO:1904923 | regulation of autophagy of mitochondrion in response to mitochondrial depolarization | $1.10 \times 10^{-3}$ |